\documentclass[prl]{revtex4}
\usepackage{graphicx}
\begin{document}

\title{Statistics of multiple-scatterer induced frequency splitting in whispering gallery microresonators and microlasers}

\author{Lina He, \c{S}ahin Kaya \"{O}zdemir, Jiangang Zhu, Monifi Faraz, Huzeyfe Y{\i}lmaz and Lan Yang}

\address{Electrical and Systems Engineering, Washington University in St. Louis, St. Louis, MO 63130, USA}

\begin{abstract}
We investigate numerically and experimentally the statistics of the changes in the amount of frequency splitting upon the adsorption of particles one-by-one into the mode volume of whispering gallery mode (WGM) microresonator and microlasers. This multiple-particle induced frequency splitting (MPIFS) statistics carries information on the size and the number of adsorbed particles into the mode volume, and it is strongly affected by two experimental parameters, namely the WGM field distribution and the positions of the particles within the mode volume. We show that the standard deviation and maximum value of the MPIFS are proportional to the polarizability of the particles, and propose a method to estimate particle size from the MPIFS if the only available data from experiments is frequency splitting.
\end{abstract}

\maketitle

\section{Introduction}\label{sec:introduction}
In optical microresonators with circular boundaries, whispering gallery modes (WGMs) traveling along the resonator perimeter are supported. Such resonators have the ability to trap light in a strongly confined area with extremely low losses \cite{Vahala_cavity,WGM1,WGM2}. They have attracted great interests in the past few years due to their high quality factors ($Q$) and small mode volumes ($V$), which enable ultra-high light intensity within the resonators \cite{Sahin_Purcell}. This makes WGM resonators promising platforms to study light-matter interactions and nonlinear optics, and to achieve high-sensitivity label-free sensing of biological and chemical materials as well as nanoparticles \cite{Vollmer_sensor,Jiangang_nature,Lina_nature,Fan_particle,Arnold_36nm,Woosung_split,Woo-hemozoin}. When the evanescent tail of the WGM field interacts with the material of interest adsorbed into the resonator mode volume, the effective optical path length traversed by the WGM field along the periphery of the resonator changes,  giving rise to changes in the spectral properties (e.g. linewidth or frequency) of the resonant mode. Such changes are monitored and recorded as the sensing signal and processed to extract information on the adsorbed material.

Eigenmodes of a WGM resonator are twofold degenerate with identical polarization and resonance frequency: clockwise (CW) and counter-clockwise (CCW) travelling wave modes (TWMs) denoted as $a_{cw}$ and $a_{ccw}$, respectively. When sub-wavelength scatterers enter the mode volume, back-scattering of the WGM field from the scatterers into the cavity mode volume couples the CW and CCW modes, lifting the degeneracy (i.e., mode splitting or frequency splitting). As a result, two new orthogonal eigenstates which are standing wave modes (SWMs) appear as superpositions (i.e., $(a_{cw}+a_{ccw})/\sqrt{2}$ and $(a_{cw}-a_{ccw})/\sqrt{2}$) of the two counter-propagating TWMs \cite{Jiangang_nature,Lina_nature, Split_Weiss,Split_Gorodetsky,Split_Deych,Yasha Yi,Split_QD}. Since light-scatterer interaction strength is proportional to the light intensity at the location of the scatterer within the mode volume, two SWMs are affected differently by the scatterers, because the intensity distributions of these two SWMs are $\pi$/2 phase-shifted from each other (i.e, phase here refers to the spatial distance between the nodes of the SWMs, and the distance between two adjacent nodes corresponds to $\pi$). This is reflected in the different resonance frequencies and linewidths of the two SWMs. When frequency splitting of the two modes is larger than their mean linewidth, the two SWMs are resolved as a doublet (two resonant modes) in the transmission spectrum \cite{Split_Mazzei,Zhu_OE2010,Split_RealSpace}.

\begin{figure}[h]
  \centering
  \includegraphics{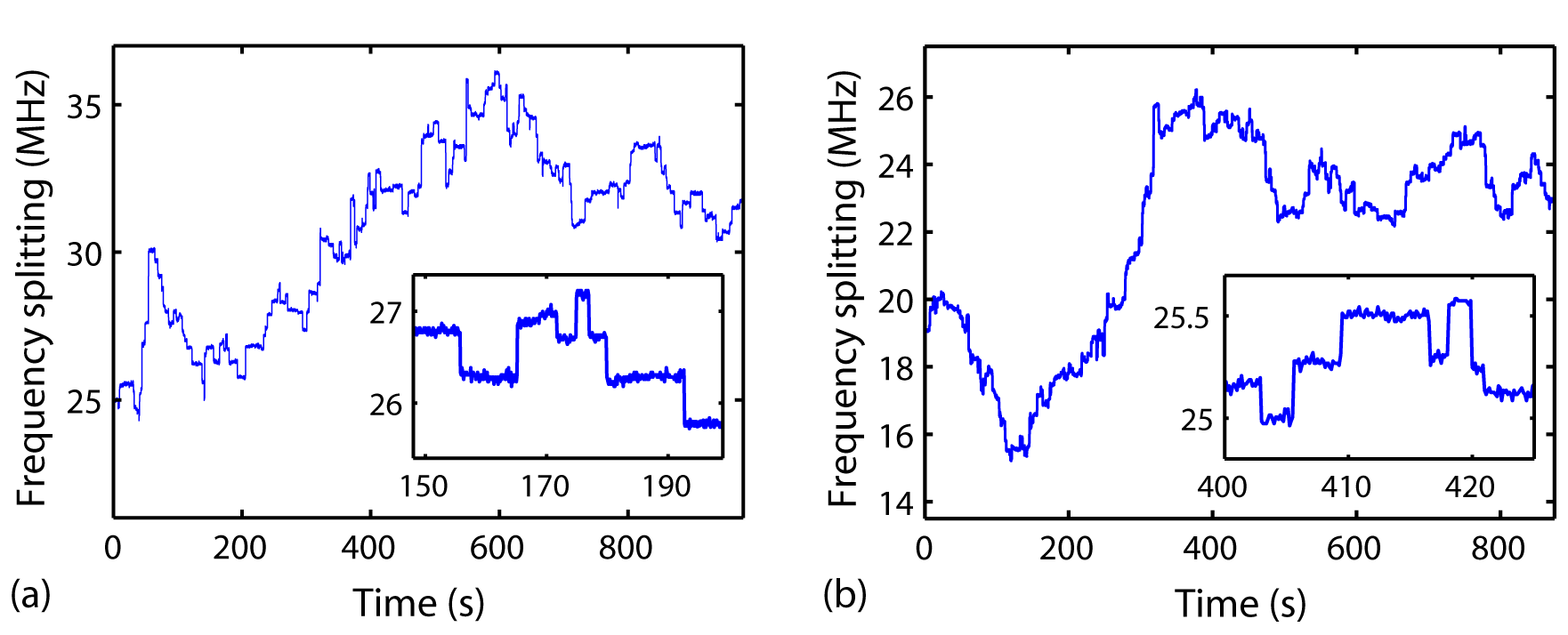}
  \caption{Experimentally obtained frequency splittings as gold nanoparticles of radius $R=30~{\rm nm}$ are continuously deposited onto two microtoroids (a) and (b). Each discrete jump indicates one binding event. There are about $200$ particles detected in each plot. Insets of (a) and (b) present close-up plots of frequency splitting observed in a short time period. Axes of the insets are the same as those of the main plots. Note that plasmonics effects do not take place here because the measurements were performed with a laser at wavelength of 1550 nm, far away from the plasmon wavelengths of the used particles.}
  \label{fig:splitting}
\end{figure}

Scatterer induced mode splitting in a WGM resonator has been used to detect and size single nanoparticles and viruses adsorbed onto the resonator \cite{Jiangang_nature,Lina_nature,Tao_Particle}. The two resonance modes forming upon splitting act as a self-referencing system \cite{Lina_SelfRef} helping to discriminate  interfering perturbations from the ones of interest and to minimize, if not eliminate, the effects of some noise sources which affect both modes on the detection and size estimation. Figure \ref{fig:splitting} shows a series of measured frequency splitting in response to adsorption of Polystyrene (PS) nanoparticles on a microtoroidal resonator. Each discrete change in frequency splitting corresponds to one particle binding event. Since WGM field distribution along the resonator surface is not uniform, the heights of discrete upward or downward steps differ for each binding event. Unlike single-side shift of resonance-shift based detection, frequency splitting either increases or decreases with binding particles \cite{Lina_nature}. As a PS particle attaches onto a silica resonator, resonances of the split modes shift to lower-frequency side (due to the higher refractive index of PS than that of air), with the shift amount determined by overlap of the particle with the field distributions of the two SWMs. If the higher-frequency mode shows a larger shift than the lower-frequency mode as shown in the lower panel of Fig. \ref{fig:splitting}(b), frequency splitting decreases leading to observed downward discrete step in Fig. \ref{fig:splitting}(a).

In real-life applications, it is not easy to control or measure the positions of individual particles on the resonator. Thus, the change in frequency splitting appears to be random. It would be interesting to investigate the statistics of frequency splitting in response to a large number of particle binding events. We demonstrate that statistical properties of multiple-particle induced frequency splitting (MPIFS) are related to the polarizability of the particle and thus can be used to extract the information of particle polarizability. We aim to answer the following question: {\it Can we estimate the number and average size of particles in the mode volume of a resonator if the statistics is provided?}  In the following parts, we first introduce a theoretical model to carry out numerical simulations on MPIFS, and then present a comparative discussion on the results of experiments and numerical simulations of MPIFS. In this study, we consider particles much smaller than the optical wavelength and thus can model nanoparticles as dipoles such that the evanescent field is uniform over each particle. For numerical simulations and experiments, we used microtoroidal resonators as the model device but the results are valid for other types of WGM optical resonators.

\section{Theoretical model}\label{sec:theory}

Theoretical models of MPIFS based on different frameworks have been investigated in previous work \cite{Chantada,Xiao_MultiParticle,Zhu_OE_MultiVirus}. In this section, we briefly introduce the model we employ for this study.

\subsection{Polarizability.}  In scatterer-induced mode splitting, the amount of mode splitting $2g$ (spectral distance between the split modes)
\begin{equation}\label{ggg1}
2g=-\frac{\alpha f^2(\textbf{r})\omega_{\rm c}}{V}
\end{equation}
and the additional dissipation $2\Gamma$ (linewidth difference of the split modes)
\begin{equation}\label{ggg2}
2\Gamma=\frac{\alpha\!^2f^2(\textbf{r})\omega_{\rm c}\!^4}{3\pi\nu^3V}
\end{equation}
are functions of the polarizability $\alpha$ \cite{Jiangang_nature,Split_Mazzei}. Their ratio
\begin{equation}\label{ggg3}
\frac{2\Gamma}{2g}=-\frac{\alpha}{3\pi}\left(\frac{2\pi n_{\rm m}}{\lambda}\right)^3
\end{equation}
allows one to estimate the polarizability from experimental data ($2g$ and $2\Gamma$ can be measured in the experiments) without the need for knowing the location of the particle on the resonator and the mode volume $V$ of the WGM \cite{Jiangang_nature}. Therefore, it is worth discussing on how to define or assign polarizability accurately to particles. In (\ref{ggg1})-(\ref{ggg3}), $f(\textbf{r})$ represents the normalized (i.e., normalized to the maximum value) WGM field magnitude at the particle position $\textbf{r}$, $\omega_{\rm c}$ is the resonance angular frequency, $n_{\rm m}$ is the refractive index of the surrounding medium, and $\nu$ is the speed of light in the surrounding medium.

The concept of polarizability arises in calculating the response of a particle in an electromagnetic field. In the simplest form where the particle is a small sphere and the field is uniform over the particle, this response can be calculated by assigning an induced dipole whose polarizability satisfies Clausius-Mossotti relation. In the electrostatic limit, $\alpha$ for a homogenous spherical particle of radius $R\ll\lambda$ (Rayleigh approximation), where $\lambda$ is the wavelength of the light, is given by
\begin{equation}
\alpha_{\rm sphere}=4\pi R^3n_{\rm m}\!^2\frac{n_{\rm p}\!^2-n_{\rm m}\!^2}{n_{\rm p}\!^2+2n_{\rm m}\!^2}
\label{alphaxxx}
\end{equation}
with $n_{\rm p}$ and $n_{\rm m}$ denoting the refractive indices of the particle and the surrounding medium, respectively. Clearly the polarizability depends on the volume and the dielectric constant of the particle, as well as the dielectric constant of the environment the particle is embedded in. At the Fr\"{o}clich frequency which minimizes $(n_{\rm p}\!^2+2n_{\rm m}\!^2)$, the polarizability $\alpha_{\rm sphere}$ and hence the induced homogenous polarization inside the particle experience a resonant enhancement associated with dipolar surface plasmons. For a non-absorbing surrounding, $n_{\rm m}\!^2$ is real, and hence the resonant condition reduces to ${\rm Re}[n_{\rm p}\!^2]=-2n_{\rm m}\!^2$. The non-vanishing imaginary part ${\rm Im}[n_{\rm p}\!^2]$ limits the magnitude of the polarizability at resonance. The dielectric constant (i.e., refractive index) of metal particles is strongly dependent on the wavelength (i.e., frequency $\omega$) of the incident light (i.e., $n_p(\omega)$), and is negative for most frequencies in the visible range. Thus it can satisfy the resonance condition.

For non-spherical particles such as ellipsoid or rod-like particles, the expression in (\ref{alphaxxx}) is no longer valid. For example, the dipolar polarizability of an ellipsoid with principal axes $a$, $b$ and $c$  should be defined along each of its principal axes  $j$, and is given as \cite{Maier,Bohren}
\begin{equation}
\alpha^{\rm (j)}_{\rm ellipsoid}=\frac{4}{3}\pi abc n_{\rm m}\!^2\frac{n_{\rm p}\!^2-n_{\rm m}\!^2}{n_{\rm p}\!^2+L_j n_{\rm m}\!^2}
\label{alphaeee}
\end{equation}
where $L_j$ denotes the geometrical depolarization factor satisfying $\sum L_j=1$. For spherical particles, since $a=b=c=R$, we have $L_1=L_2=L_3=1/3$, implying that a spherical particle has the same dipolar polarizability along all directions. For spheroidal particles with $a=b$, on the other hand, we have $L_1=L_2$ and the depolarization effects depend on $c/a$.

For particles where $R\ll\lambda$ is not fully satisfied (e.g., $R>25~{\rm nm}$ in the visible wavelength), the particle does not experience a uniform field over its volume, and the electrostatic limit cannot be justified. In this case, retardation effects must be accounted for \cite{Kuwata,Abajo,Russell}. Retardation effects take place in such particles because opposite charges in the induced dipole mode become largely separated - approximately one particle diameter- such that one end of the particle feels the changes in the other end with a phase delay due to the finite speed of light. Thus the period of the oscillations increases to accommodate this phase delay. Hence, a reduction in the depolarization field and mode energy is seen. This phase difference over the particle volume is especially significant for particles of larger size (e.g., $R>50~{\rm nm}$ illuminated with visible light). Consequently, the expression for  $\alpha$ given in (\ref{alphaxxx}) should be modified to take the retardation effects into consideration as suggested by Kuwata {\it et al.} \cite{Kuwata} for particles of arbitrary shapes. Indeed, for large spherical particles, retardation effects can be seen in the multipole extension of the fields and the corresponding Mie coefficients. Retardation effects in large particles are manifested as red-shift of the plasmon resonance, appearance of higher multipoles such as quadrupoles in addition to dipoles, and radiative loss as the particle size increases. For small nanoparticles (e.g., $R<25~{\rm nm}$) in the electrostatic limit, on the other hand, plasmon frequency is insensitive to particle size.

Polarization of the light plays an important role in the light-nanoparticle interactions. The knowledge on the polarization properties of the light is crucial, in particular, for non-spherical nanoparticles, because the induced polarization depends on the component of the electric field along each of the principal axes. In contrast to spherical particles where the diameter is the same along different polarizations (e.g., transverse electric (TE) or transverse magnetic (TM)) of the incident light field, spheroidal and rod-like particles have different diameters along different light polarizations \cite{Teraoka,Noto}. Therefore, for such non-spherical particles, different polarizations of light field will experience difference polarizabilities and lead to different dipole moments, even for small particles satisfying the Rayleigh limit \cite{Woo-hemozoin,Maier,Bohren,Abajo}. Moreover, higher multipolar orders (i.e., the first order corresponds to small particle limit: dipolar approximation) which should be considered for larger particles are polarization dependent and should be taken into account. Therefore, to accurately assign polarizability to a detected particle of unknown shape, one should look at the response by varying the polarization of the light field.

For the specific system we consider here, that is the interaction of the WGM field with a nanoparticle, the interacting field is an evanescent field and is not homogenous over the volume of large particles. Therefore, the electrostatic approximation may not hold anymore and the multipole expansion of the polarizability and the Mie coefficients should be considered \cite{Quinten,Chaumet}. It is well-known that higher multipolar orders contribute to the scattering and absorption more in the case of evanescent fields than of plane waves, and more for larger particles than the smaller particles (i.e., Mie coefficients of higher orders decrease very rapidly for small particles, thus their contribution is not significant for smaller particles) \cite{Quinten}. Moreover, these higher orders are polarization dependent resulting in polarization dependent scattering and absorption cross-sections which are larger for TM-polarized light \cite{Bohren}.

\begin{figure}[]
  \centering
  \includegraphics{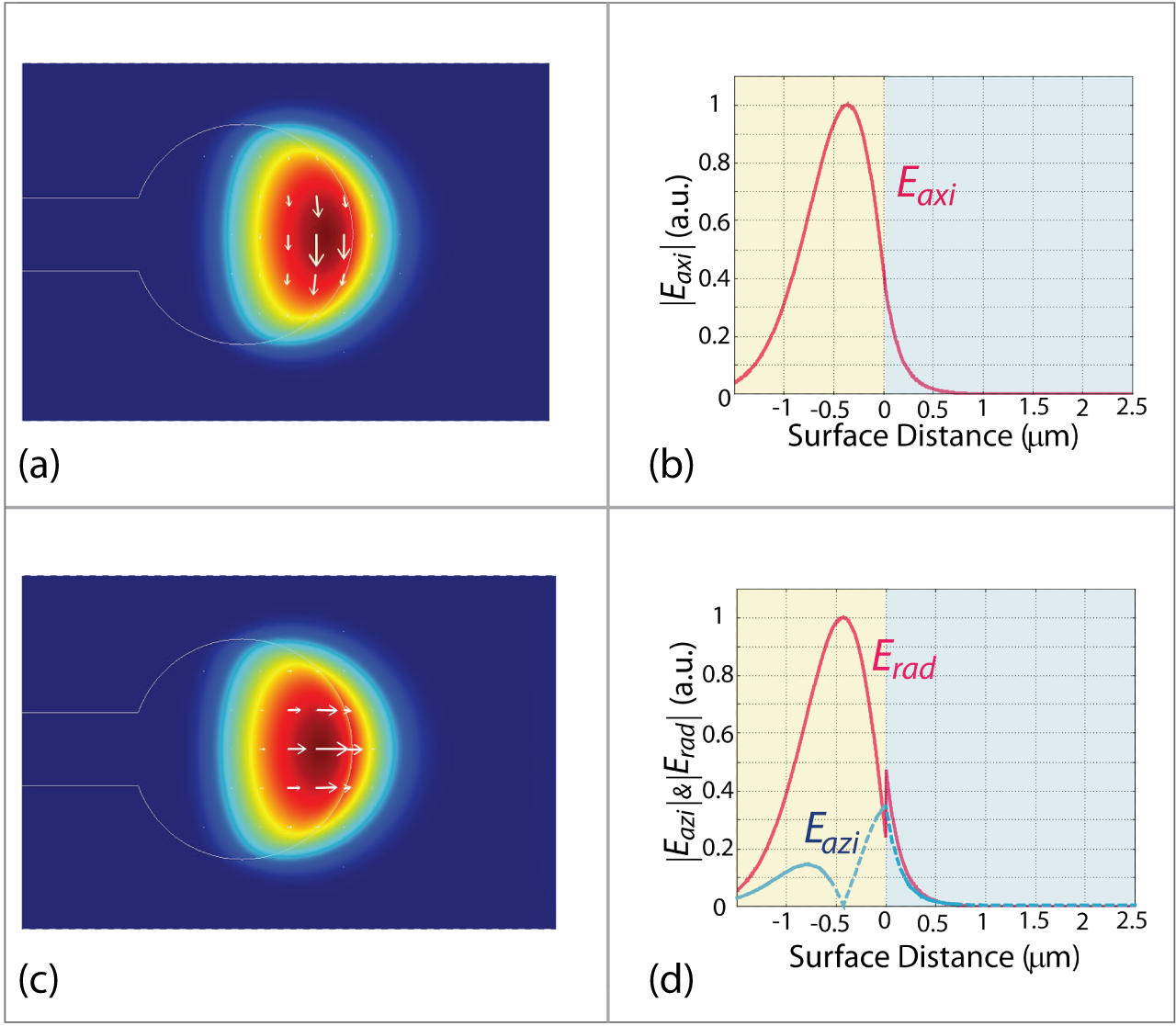}
  \caption{Results of numerical simulation carried out by COMSOL showing the electric field distributions of TE (a,b) and TM (c,d) modes in a microtoroid resonator in the 1550nm band. The white arrows denote the direction of electric field.}
  \label{fig:fielddist}
\end{figure}

In WGM resonators, the electric field component of TE modes is axial with no radial component whereas the electric field component of TM modes is predominantly in the radial direction with a relatively small azimuthal component (see Fig. \ref{fig:fielddist}). In addition, the electric field of TE modes is continuous along the boundary between the resonator and the surrounding in contrast to a discontinuity experienced by the radial component of the TM electric field. This difference in TE and TM modes will certainly modify the scattering and absorption properties of a particle placed within the WGM field, and the contribution of higher multipolar orders will be significantly higher for the TM mode than the TE mode leading to higher scattering and absorption cross-sections for the TM mode. As pointed out by Chantada {\it et al.} \cite{Chantada}, scattering of WGM field by particles may lead to resonance broadening which scales with the scattering cross-section of the particle and the polarization of the field for small particles. This type of scaling for linewidth broadening is difficult for large particles because the process is a mixture of scattering from the particle and the propagation of light within the particle that can couple back into the resonator.

In this study, we consider small spherical particles such as polystyrene (PS) nanoparticles $R\leq50~{\rm nm}$, sodium chloride (NaCl) nanoparticles of $R=25~{\rm nm}$ and $R=30~{\rm nm}$, and gold (Au) nanoparticles of $R=15~{\rm nm}$ and $R=25~{\rm nm}$. The wavelength $\lambda$ of the WGM is in the 1550 nm band, which is far from the plasmon resonances of the gold nanoparticles. Thus we do not consider plasmonic enhancement of scattering and absorption. The field is evanescent outside the resonator with a characteristic length of the order of the wavelength, hence the particles are completely within the field, although the field is not uniform over their volumes. A consequence of the polarization dependent scattering cross-section of the particles in the evanescent WGM field is that TM modes are more sensitive to particle-induced perturbations and enable the detection of smaller particles than TE modes. In our experiments, the motivation has been the detection of smaller particles. Therefore, by fiber polarization controllers we choose the polarization of the input light which allows us to detect the smallest particles possible. The polarization is kept fixed throughout the study. Under these conditions, we can also safely neglect retardation effects caused by geometric depolarization and use the polarizability expression in (\ref{alphaxxx}).

\subsection{General Model for Scatterer-Induced Mode Splitting} 
\begin{figure}[]
  \centering
  \includegraphics{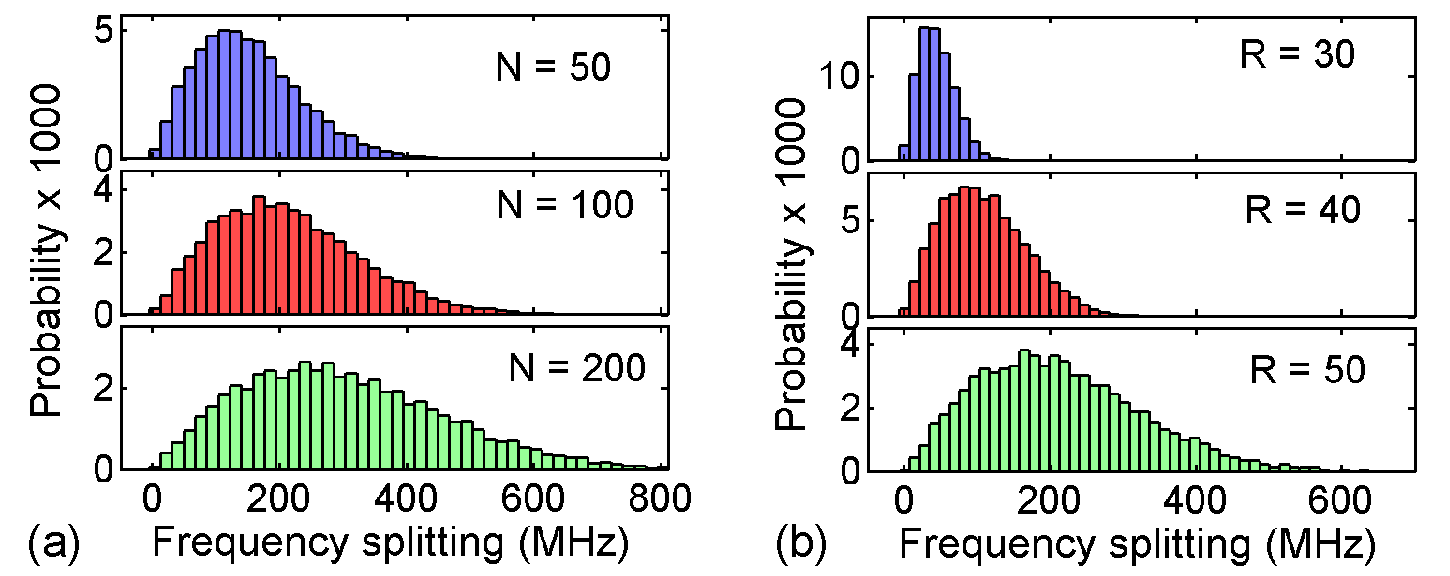}
  \caption{Distributions of frequency splitting induced by PS nanoparticles of different number $N$ (a) and size $R$ (b) in the mode volume of a microtoroid without initial mode splitting. We set  $R$ = 50 nm in (a) and $N$ = 100 in (b). Particle radii labeled in (b) have unit of nanometer. The distribution in each panel is obtained by 10000 repeated trials.}
  \label{fig:hist_splitting}
\end{figure}

The simplest case is one single particle ($N=1$) in the resonator mode volume. Figure \ref{fig:hist_splitting}(a) depicts the field distributions of the formed orthogonal SWMs. The particle locates at the node of one of the SWMs and the anti-node of the other. With more particles adsorbing onto the resonator, the spatial field distributions of the SWMs are perturbed and redistributed to maximize the coupling strength between CW and CCW modes \cite{Zhu_OE_MultiVirus}. Spatial distributions of the two SWMs depend on polarizabilities and positions of all attached particles. As a result, both SWMs experience resonance shift and linewidth broadening with the amounts determined by how much their fields overlap with the physical locations of the particles, i.e., the mode whose anti-node is closer to the particles is affected more significantly than the mode which has more particles at its node. Figure \ref{fig:hist_splitting}(b) shows the field distributions of the SWMs when a second particle ($N=2$) binds onto the resonator.

To study the effects of mode splitting quantitatively, we consider $N$ particles entering the resonator mode volume. Resonance shift and linewidth broadening of two SWMs are given by the combined effect of all individual particles whose contribution is determined by its position relative to the SWM distributions. This can be explained by the fact that the strength of the interaction between the light and the scatterer, which determines the amount of resonance shift and linewidth broadening, is proportional to the field intensity at the particle location. Here, we denote the resonance frequency and linewidth of the initial (i.e., before the adsorption of the particle) degenerate mode as ($\omega_0$,$\gamma_0$) and of the split modes after adsorption of $N$ particles as ($\omega^{(1)}_{\rm N}$,$\gamma^{(1)}_{\rm N}$) and ($\omega^{(2)}_{\rm N}$,$\gamma^{(2)}_{\rm N}$) with the superscript describing two SWMs and the subscript representing the $N$-th particle. We define $\phi_{\rm N}$ as the spatial phase difference between the first particle and the anti-node of one SWM, and $\psi_{\rm i}$ as spatial phase distance between the $i$-th particle and the first particle. Frequency shift ($\Delta \omega^{(1)}_{\rm N}$,$\Delta \omega^{(2)}_{\rm N}$) and linewidth broadening ($\Delta \gamma^{(1)}_{\rm N}$,$\Delta \gamma^{(2)}_{\rm N}$) of two SWMs after $N$ particles enter the mode volume are written as \cite{Zhu_OE_MultiVirus}
\begin{eqnarray}\label{Eq_shift}
\Delta \omega^{(1)}_{\rm N}\equiv\omega^{(1)}_{\rm N}-\omega_{0}= \sum^{\rm N}_{\rm i=1}2g_{\rm i}\cos^2(\phi_{\rm N}-\psi_{\rm i}),
\end{eqnarray}
\begin{eqnarray}\label{Eq_shifta}
\Delta \omega^{(2)}_{\rm N}\equiv\omega^{(2)}_{\rm N}-\omega_{0}=\sum^{\rm N}_{\rm i=1}2g_{\rm i}\sin^2(\phi_{\rm N}-\psi_{\rm i})
\end{eqnarray}
and
\begin{eqnarray}
\Delta \gamma^{(1)}_{\rm N}\equiv\gamma^{(1)}_{\rm N}-\gamma_{0}=\sum^{\rm N}_{\rm i=1}2\Gamma_{\rm i}\cos^2(\phi_{\rm N}-\psi_{\rm i})
\label{Eq_linewidth}
\end{eqnarray}
\begin{eqnarray}
\Delta \gamma^{(2)}_{\rm N}\equiv\gamma^{(2)}_{\rm N}-\gamma_{0}=\sum^{\rm N}_{\rm i=1}2\Gamma_{\rm i}\sin^2(\phi_{\rm N}-\psi_{\rm i})
\label{Eq_linewidtha}
\end{eqnarray}
where $2g_{\rm i}=-\alpha_{\rm i}f^2(\textbf{r})\omega_{\rm c}/V$ and $2\Gamma_{\rm i}=\alpha_{\rm i}\!^2f^2(\textbf{r})\omega_{\rm c}\!^4/(3\pi\nu^3V)$ correspond to resonance shift and linewidth broadening if the $i$-th particle is the only adsorbed particle, $\alpha_i$ is the polarizability of the $i$-th particle,  $f(\textbf{r})$ represents the normalized (i.e., normalized to the maximum value) WGM field magnitude at the particle position $\textbf{r}$, $\omega_{\rm c}$ is the resonance angular frequency, $\nu$ is the speed of light in the surrounding medium, and $V$ is the WGM mode volume.

Frequency splitting after the adsorption of $N$ particles in the resonator mode volume is found from (\ref{Eq_shift}) and (\ref{Eq_shifta}) as
\begin{equation}
S_{\rm N}\equiv\left|\Delta\omega^1_{\rm N}-\Delta\omega^2_{\rm N}\right|=\left|\sum^{\rm N}_{i=1}2g_{\rm i}\cos(2\phi_{\rm N}-2\psi_{\rm i})\right|
\label{Eq_splitting}
\end{equation}
which takes its maximum when one of the split modes is maximally shifted from the degenerate mode while the other split mode experiences a minimal shift. Recall that the resonance condition for the WGM resonator requires that an integer multiple of the light wavelength is equal to the optical path length that light travels within the resonator. A change in the optical path length shifts the resonance wavelength. Thus, the shifts of the split modes are related to the change in the optical path lengths they propagate within the resonator, i.e., increase in the amount of splitting corresponds to increase in their optical path difference. For the split mode which experiences the maximal (minimal) shift with respect to the degenerate mode, the optical path length should change maximally (minimally). This is consistent with the modern interpretation of Fermat's principle that the optical length of the path followed by light between two fixed points is an extremum. Thus, we can analyze the shifts of the split modes and determine the conditions for their maximization (or minimization) by taking the derivative of the expressions given in (\ref{Eq_shift}) and (\ref{Eq_shifta}) with respect to $\phi_{\rm N}$. Taking the first derivatives and arranging the resultant expressions using trigonometric identities, we arrive at
\begin{eqnarray}
\frac{\partial\Delta \omega^{(1)}_{\rm N}}{\partial\phi_{\rm N}}&&=-2\sum^{\rm N}_{\rm i=1}g_{\rm i}\sin(2\phi_{\rm N}-2\psi_{\rm i})\\ \nonumber
&&=2\cos(2\phi_{\rm N})\sum^{\rm N}_{\rm i=1}g_{\rm i}\sin(2\psi_{\rm i})-2\sin(2\phi_{\rm N})\sum^{\rm N}_{\rm i=1}g_{\rm i}\cos(2\psi_{\rm i})\\ \nonumber
&&=-\frac{\partial\Delta \omega^{(2)}_{\rm N}}{\partial\phi_{\rm N}}.
\label{Eq_splittinga}
\end{eqnarray}
Setting $\frac{\partial\Delta \omega^{(1)}_{\rm N}}{\partial\phi_{\rm N}}=0$ and $\frac{\partial\Delta \omega^{(2)}_{\rm N}}{\partial\phi_{\rm N}}=0$, we find that the functions in (\ref{Eq_shift}) and (\ref{Eq_shifta}) have their extrema (critical points) at the values of $\phi_{\rm N}$ satisfying
\begin{equation}
\tan(2\phi_{\rm N})=\frac{\sum^{\rm N}_{\rm i=1}g_{\rm i}\sin(2\psi_{\rm i})}{\sum^{\rm N}_{\rm i=1}g_{\rm i}\cos(2\psi_{\rm i})}.
\label{Eq_phi}
\end{equation}
Taking the second derivatives of (\ref{Eq_shift}) and (\ref{Eq_shifta}) and inserting the expression in (\ref{Eq_phi}), we arrive at
\begin{eqnarray}
\frac{\partial^2\Delta \omega^{(1)}_{\rm N}}{\partial\phi_{\rm N}^2}&&=-4\sec(2\phi_{\rm N}\sum^{\rm N}_{\rm i=1}g_{\rm i}\cos(2\psi_{\rm i})\\ \nonumber
&&=-\frac{\partial^2\Delta \omega^{(2)}_{\rm N}}{\partial\phi_{\rm N}^2}.
\label{Eq_splittingb}
\end{eqnarray}
It is then clear that for $\sum^{\rm N}_{\rm i=1}g_{\rm i}\cos(2\psi_{\rm i})\neq 0$, extrema corresponding to minima (maxima) for $\Delta \omega^{(1)}_{\rm N}$ (i.e., shift of the lower frequency split mode) are maxima (minima) for $\Delta \omega^{(2)}_{\rm N}$ (i.e., shift of the higher frequency split mode) and vice versa. Thus, one of the SWMs has maximum resonance shift with respect to the degenerate mode whereas the other one shows minimal shift, giving rise to maximum splitting.  For $\sum^{\rm N}_{\rm i=1}g_{\rm i}\cos(2\psi_{\rm i})= 0$, we find from (\ref{Eq_phi}) that $\tan(2\phi_{\rm N})=\mp \infty$ implying $2\phi_{\rm N}=m\pi+\pi/2$ and hence $\cos(2\phi_{\rm N})=0$ and $\sin(2\phi_{\rm N})=\mp 1$. Substituting this into (\ref{Eq_splitting}), we obtain $S_{\rm N}=\left|\sum^{\rm N}_{i=1}2g_{\rm i}\sin(2\psi_{\rm i})\right|$. For each new particle entering the mode volume, the distributions of the two SWMs rotate so as to ensure that $\phi_{\rm N}$ satisfies (\ref{Eq_phi}). This takes place at each binding event leading to changes in mode-splitting spectrum which enable detection and size measurement of binding particles.

Single-shot size measurement of individual nanoparticles and viruses have been demonstrated by tracking the changes in the resonance frequencies and linewidths of the split modes in a passive microtoroid \cite{Jiangang_nature,Zhu_OE_MultiVirus}. Since the difference of the linewidths and the resonance frequencies of the split modes both scale as $f^2(\textbf{r})/V$, their ratio $g/\Gamma$ becomes independent of the particle position and the resonator mode volume, enabling estimation of the size of each detected particle without knowing its position in the mode volume and the size of the resonator mode volume. It is apparent that this size measurement method requires that the changes in the resonance frequencies and the linewidths are measured with high accuracy. However, as the size $R$ of the particle decreases, the changes in the linewidths become too small to be measured with high accuracy as the parameter $\Gamma$ scales as $R^6$. Since the parameter $g$ is proportional to $R^3$, there is a region of $R$ below which resonance frequency changes can be measured but the changes in linewidths cannot be resolved. Thus, size measurement either cannot be made or can be erroneous although the detection of the particle is still possible.

In a recent work \cite{Lina_nature}, we have demonstrated that mode splitting in a WGM microlaser \cite{Lina_Rev} enables enhanced detection sensitivity capable of detecting smaller particles beyond the reach of passive resonators employing mode splitting. Each of the split modes is a lasing line which when detected at a photodetector gives rise to a beat signal whose frequency is the same as the amount of mode splitting. This provides an easy way to detect splitting and hence monitor continuously the binding of nanoparticles onto the microlaser. However, it is almost impossible to perform single-shot measurement of linewidth difference of the split lasing lines as the laser linewidths are very small. Thus, although smaller particles can be detected by monitoring the beat frequency, single size measurement cannot be done due to the lack of information on the linewidth difference. In such cases, where changes in frequencies are detected, we can extract an average size for the ensemble of adsorbed particles. It is thereby important to investigate how the splitting spectra are affected as multiple particles are entering the resonator mode volume.

In Sections \ref{sec:freq splitting} and \ref{sec:spliting changes}, we study numerically multiple-particle induced frequency splitting in microtoroids as a function of particle number and size when taking into account particle positions in the mode volume. In all numerical simulations, we consider the WGM with distribution shown in the lower panel of Fig. \ref{fig:fielddist} and a wavelength of $\lambda=1550\ \rm {nm}$. The maximum field is $f_{\rm {max}} = 0.36$ and the mode volume is $V = 280\ \rm{\mu m}^3$. We assume particles enter the mode volume sequentially and uniformly, i.e., $\psi$ has uniform distribution from 0 to $\pi$ and $\theta$ has uniform distribution from 0 to $2\pi /3$. These assumptions are reasonable for our measurements in which particles land on different positions of the resonator with equal possibility. Moreover, in the numerical simulations we assume that all the nanoparticles in the ensemble are spherical and have the same size and polarizability. However, the analysis can be extended to ensembles where the size or the polarizability distribution of the nanoparticles follows a certain statistical distribution such as Gaussian or Poissonian.

\section{Multiple-particle induced frequency splitting $S_{\rm N}$}\label{sec:freq splitting}

As indicated by Eq. (\ref{Eq_splitting}), the frequency splitting $S_{\rm N}$ changes with each particle entering the resonator mode volume. Figure \ref{fig:splitting} shows the experimentally obtained $S_{\rm N}$ as a function of time as gold nanoparticles are consecutively deposited onto a microtoroid. Measurement details are explained in \cite{Lina_nature}. In Fig. \ref{fig:splitting}, we see that $S_{\rm N}$ either decreases or increases with each particle and there is no observed trend which can relate $S_{\rm N}$ to the number of adsorbed particles, indicating that $S_{\rm N}$ varies with particle position on the resonator. It is interesting to look at the $N$-particle induced $S_{\rm N}$ statistically which can be obtained by a large number of repeated trials.

\begin{figure}[]
  \centering
  \includegraphics{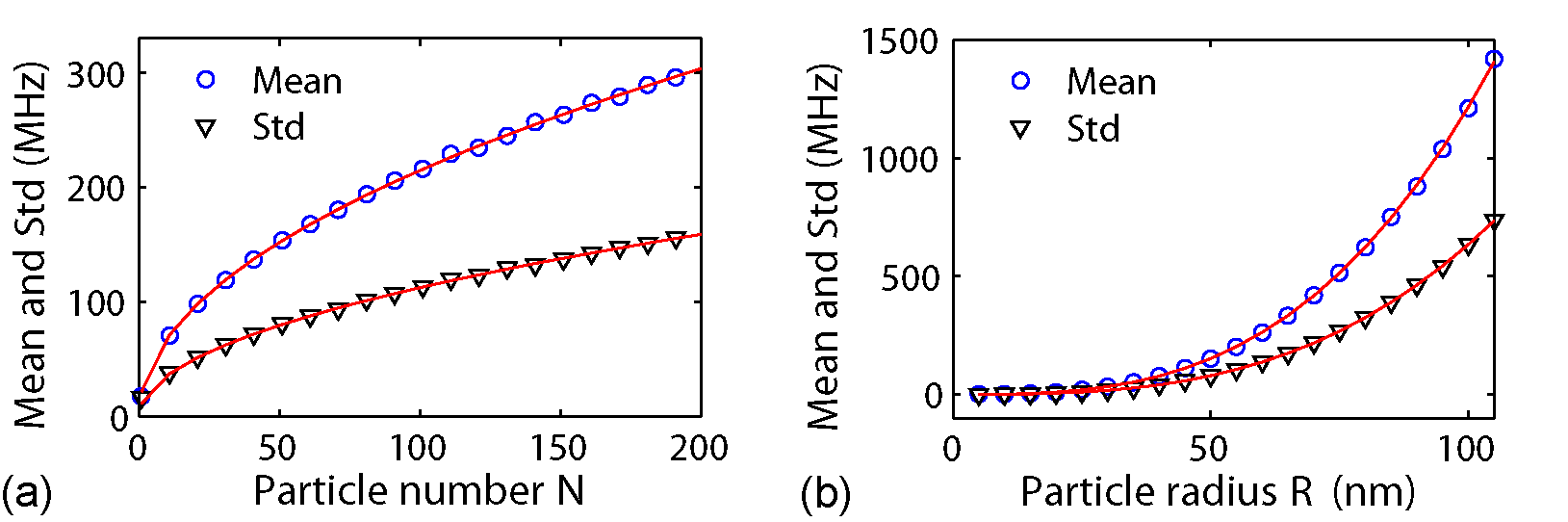}
  \caption{Mean (blue circles) and standard deviation (black triangles) of the frequency splitting induced by PS particles for different particle number $N$ (a) and radius $R$ (b). We set $R$ = 50 nm in (a) and $N$ = 50 in (b). For each $N$ and $R$, $S_{\rm N}$ is calculated 10000 times to obtain $S_{\rm N}\!^{\mu}$ and $S_{\rm N}\!^{\sigma}$. Red curves in (a) are linear fitting to $\sqrt{N}$ and in (b) are linear fitting to $R^3$.}
  \label{fig:splitting_N_R}
\end{figure}

In numerical simulations, $N$ particles with radius $R$ are randomly deposited in the mode volume of a resonator, and the induced frequency splitting is calculated using Eqs. (\ref{Eq_shift}),(\ref{Eq_phi}) and (\ref{Eq_splitting}). This process is repeated for $10000$ times to obtain a statistically significant distribution of $S_{\rm N}$. Distributions of $S_{\rm N}$ obtained for various values of $N$ and $R$ are given in Fig. \ref{fig:hist_splitting}, which depicts that the distribution of frequency splitting becomes broader (i.e., standard deviation increases) and mean frequency splitting shifts to higher values as the size of the particles and the number of deposited particles increase. The mean $S_{\rm N}\!^{\mu}$ and the standard deviation $S_{\rm N}\!^{\sigma}$ of the distributions of $S_{\rm N}$ are given in Fig. \ref{fig:splitting_N_R} as a function of $N$ and $R$, from which the curve fittings reveal $S_{\rm N}\!^{\mu}\propto\alpha\sqrt{N}$ and $S_{\rm N}\!^{\sigma}\propto\alpha\sqrt{N}$ (i.e., $S_{\rm N}\!^{\sigma}\propto R^3\sqrt{N}$). These agrees well with the results in Ref. \cite{Chantada}. The coefficients of the linear relations are determined by the WGM field distribution $f$ and the particle positions $\psi$ and $\theta$. It should be pointed out that for each single set of experiment realization, the frequency splitting is random and does not necessarily follow the curves in Fig. \ref{fig:splitting_N_R}.

\section{Particle-induced changes in frequency splitting $\Delta S_N$}
\label{sec:spliting changes}

\begin{figure}[]
  \centering
  \includegraphics{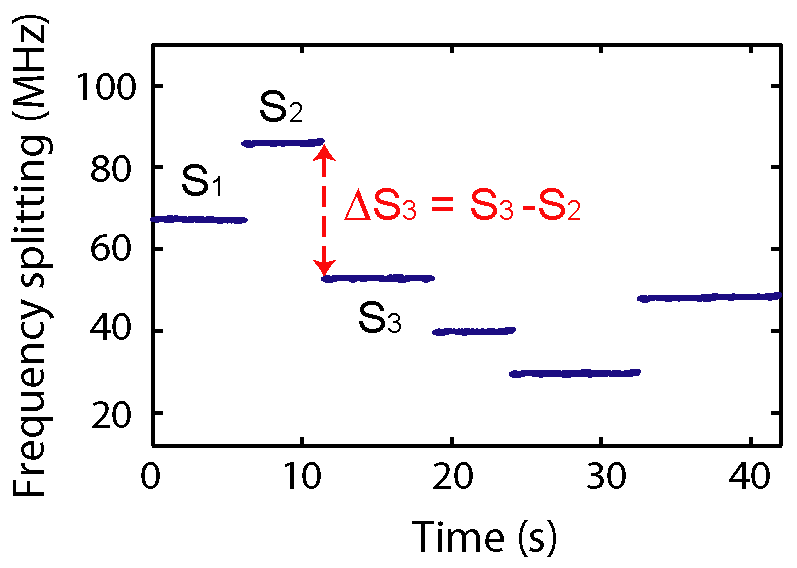}
  \caption{Definition of change in frequency splitting corresponding to single particle binding event.}
  \label{fig:Split_Change}
\end{figure}

In this section, we study the change in frequency splitting in response to adsorption of individual particles. In other words, we study the statistics of the amount of change in the frequency splitting of the WGM of interest upon binding of single particles in order to estimate the size or polarizability of the particles and the number of adsorbed particles. Here, we define the change of frequency splitting induced by the binding of the $i$-th particle as $\Delta S_{\rm i}=S_{\rm i}-S_{\rm {i-1}}$ (Fig. \ref{fig:Split_Change}).

\begin{figure}[]
  \centering
  \includegraphics{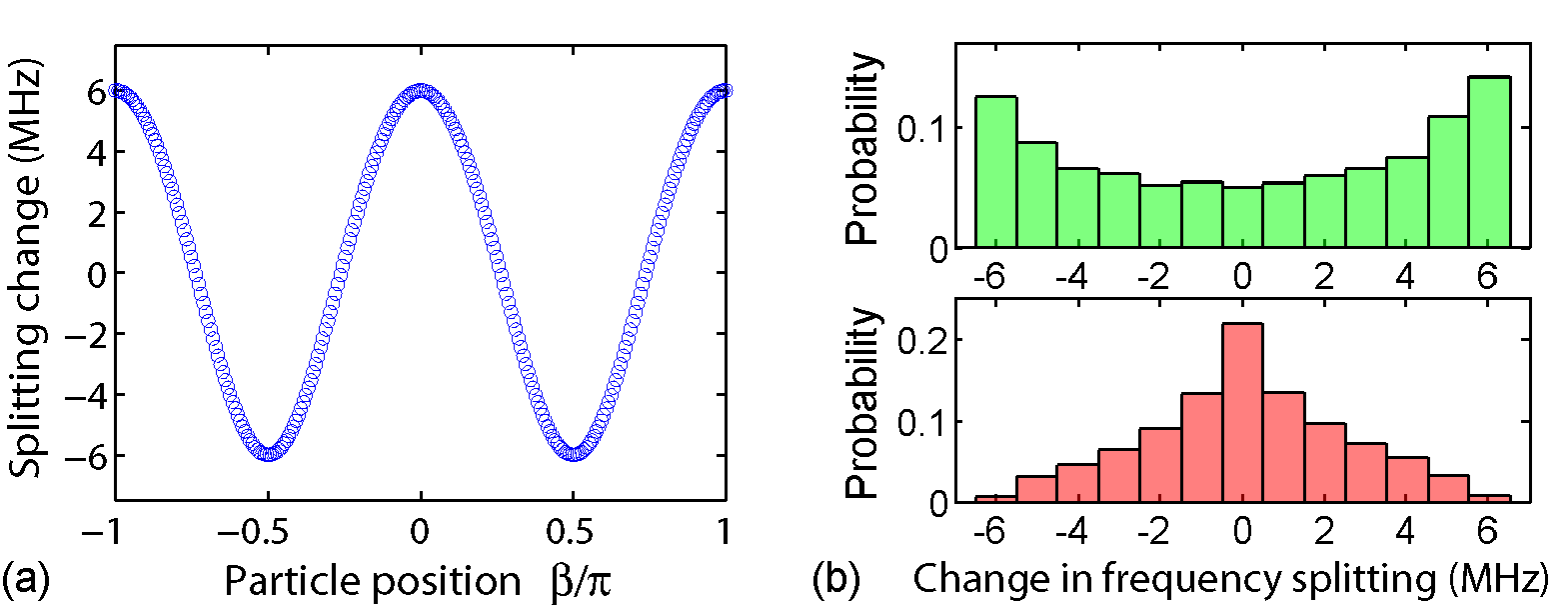}
  \caption{Simulation results showing the change in frequency splitting by a second binding particle. We set $S_1=$ 20 MHz for calculations. (a) Frequency splitting change $\Delta S_2$ as a function of particle position $\psi$ when $2g_2=$ 6 MHz. It is seen that $\Delta S_2$ varies from $-6$ MHz to 6 MHz. (b) Histograms of $\Delta S_2$ for different distributions of $2g_2$ and $\psi$. Upper panel: $2g_2$ is set as 6 MHz, while $\psi$ has uniform distribution U$\sim(0,\pi)$. Lower panel: $2g_2$ has uniform distribution U$\sim$(0, 6 MHz) due to the uncertain position $\theta$, while $\psi$ has a uniform distribution U$\sim(0,\pi)$. The distributions are obtained from 10000 trials.}
  \label{fig:TwoParticle_StepSplit}
\end{figure}

As discussed in Section \ref{sec:theory}, for an ideal resonator without particles, the CW and CCW modes are degenerate without frequency splitting, i.e., $S_0 = 0$. As the first particle enters the resonator mode volume, we have $\psi_1=0$ and $\phi_1=0$. Frequency splitting is then given as $S_1=\left|2g_1\right|$ ($\Delta S_1=\left|2g_1\right|$) which is proportional to particle polarizability, and is related to the particle position $\theta$ in the mode volume. When a second particle enters the mode volume, the induced frequency splitting is written as
\begin{equation}
S_2=\left|2g_1\cos(2\phi_2-2\psi_1)+2g_2\cos(2\phi_2-2\psi_2)\right|
\label{Eq:S2}
\end{equation}
with $\psi_1=0$, and $\phi_2$ determined by
\begin{equation}
\tan(2\phi_2)=\frac{g_2\sin(2\psi_2)}{g_1+g_2\cos(2\psi_2)}.
\label{Eq:phi2}
\end{equation}
Inserting (\ref{Eq:phi2}) in (\ref{Eq:S2}), we obtain
\begin{equation}
S_2=2\sqrt{g_1\!^2+g_2\!^2+2g_1g_2\cos(2\psi_2)}
\label{Eq:S2_2}
\end{equation}
which shows the dependence of $S_2$ on $g_1$, $g_2$, and the distance between the two particles (i.e., $\psi=\left|\psi_1-\psi_2\right|=\left|\psi_2\right|$ since $\psi_1=0$). In the special case of $g_1 = g_2 = g$, we find $\phi_2=\psi_2/2$ indicating that one of the SWMs has its anti-node at the midpoint between the two particles. Frequency splitting in this case is $S_2=\left|4g\cos(\psi)\right|$ which is maximized (minimized) for $\psi=m\pi$ ($\psi=m\pi+\pi/2$) for an arbitrary integer $m$. Moreover, (\ref{Eq:S2_2}) implies that $S_2$ varies in the range between  $\left|2g_1-2g_2\right|$ and $\left|2g_1+2g_2\right|$  for varying values of $\psi$. Thus change in the frequency splitting $\Delta S_2=S_2-S_1$ in response to the second particle position is in the range from $-\left|2g_2\right|$ to $\left|2g_2\right|$ for a large value of $2\left|g_1\right|$. This is demonstrated by the simulation result given in Fig. \ref{fig:TwoParticle_StepSplit}(a). In practice, $2g_2$ is unpredictable because of the unknown $\theta$. Therefore, uncertainty of the second particle position affects the frequency splitting through both $2g_2$ and $\psi$. In Fig. \ref{fig:TwoParticle_StepSplit}(b), we give the distribution of $\Delta S_2$ for different values of $2g_2$ and $\psi$. Statistically, for a large number of repeated tests, $\Delta S_2$ follows some distribution determined by the distributions of $2g_2$ and $\psi$. If the two particles are identical, we find from Eq. (\ref{Eq:S2_2}) that $\Delta S_2$ is proportional to the particle polarizability $\alpha$. These discussions can be readily extended to more particles, in which case the established frequency splitting is analogous to $S_1$ and the new coming particle is analogous to the second particle. In general, as the $i$-th particle enters the resonator mode volume, frequency splitting could either increase or decrease, and the amount of change depends on the position of the $i$-th particle on the resonator with its maximum possible value equal to $2g_i$.

\begin{figure}[]
  \centering
  \includegraphics{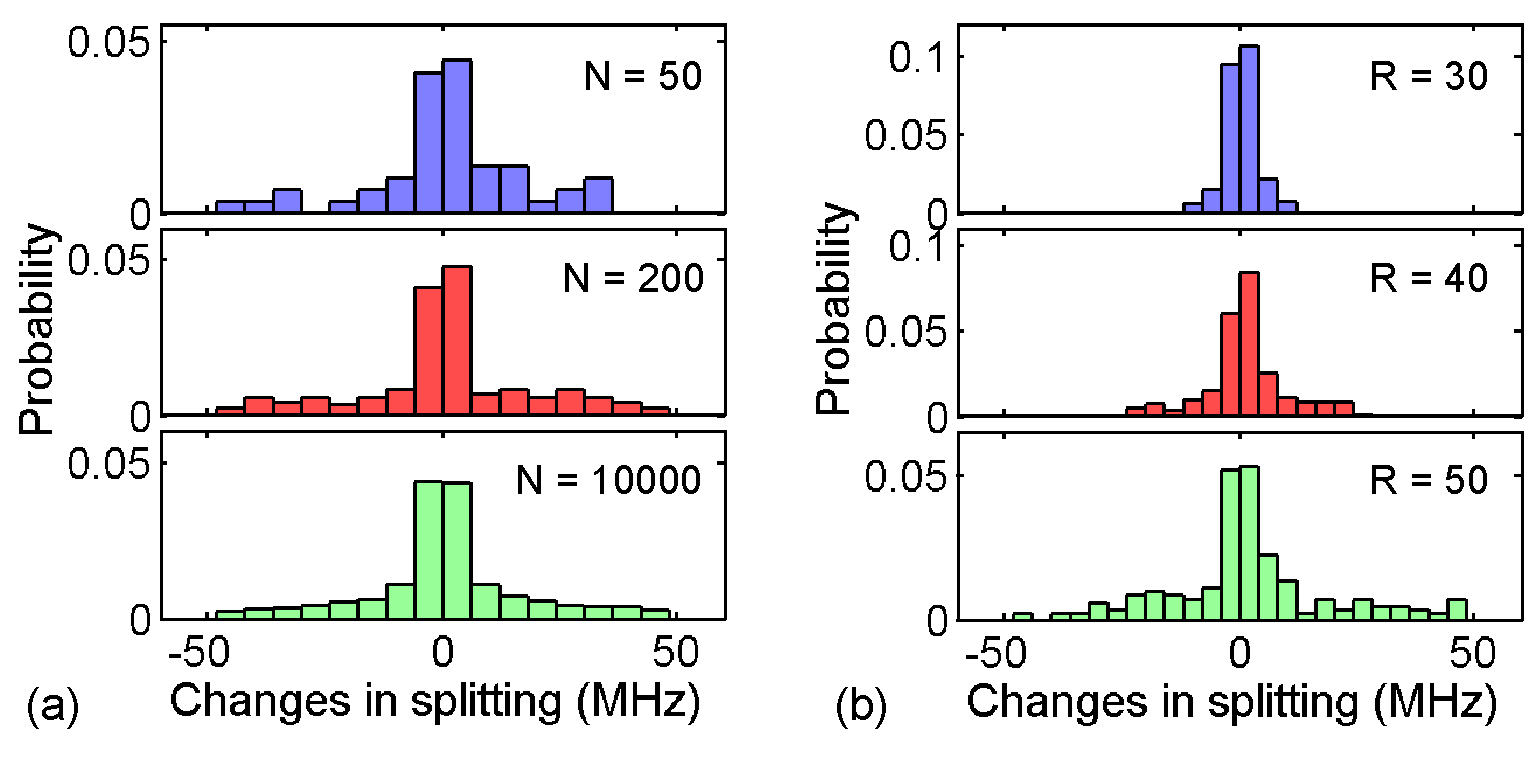}
  \caption{Histograms of splitting changes induced by consecutive deposition of PS particles onto a microtoroid. Here $R$ = 50 nm in (a) and $N$ = 200 in (b). Particle radii labeled in (b) have unit of nanometer. Each histogram corresponds to one set of experiment: continuously deposit $N$ particles of radius $R$ on a resonator, record the corresponding changes in frequency splitting, and plot the histogram of those changes. Mean and standard deviation of these distributions are listed in the form of mean/standard deviation with unit of MHz: (a) 0.7063/16.7950, 0.8465/18.2432, 0.4465/17.4071 from top to bottom; (b) 0.2421/3.3724, 1.0912/8.5915, 1.6585/17.5447 from top to bottom.}
  \label{fig:step_N_R}
\end{figure}

$N$ particles entering the mode volume of a resonator one-by-one consecutively lead to $N$ discrete changes in frequency splitting $\Delta S_{\rm i}$, $i=1\ldots N$. Figure \ref{fig:step_N_R} presents histograms of these changes ($\Delta S_1\sim\Delta S_{\rm N}$) for various $N$ and $R$. Theoretically, for a large enough number of particle binding events, the histogram of splitting changes is symmetric around zero as shown in the bottom panel of Fig. \ref{fig:step_N_R}(a). The maximum possible change in the histogram equals to the maximum value of $2g$, i.e., $\Delta S^{\rm max}=\alpha f_{\rm max}\!^2\omega_c/V$, which is achieved at $\theta = 0$. The standard deviation $\Delta S^{\sigma}$ of the histogram is proportional to $\alpha$ with the coefficient determined by the distribution of particle positions. These are demonstrated by the negligible impact of particle number $N$ on the distribution of $\Delta S$ when $N$ is sufficiently large (Fig. \ref{fig:step_N_R}(a)), and by the broader distribution of $\Delta S$ with increasing $R$ (Fig. \ref{fig:step_N_R}(b)). The dependence of $\Delta S^{\sigma}$ and $\Delta S^{\rm max}$ on $\alpha$ can be used to extract the information of particle polarizability and thus particle size \cite{Lina_nature}.

\begin{figure}[]
  \centering
  \includegraphics{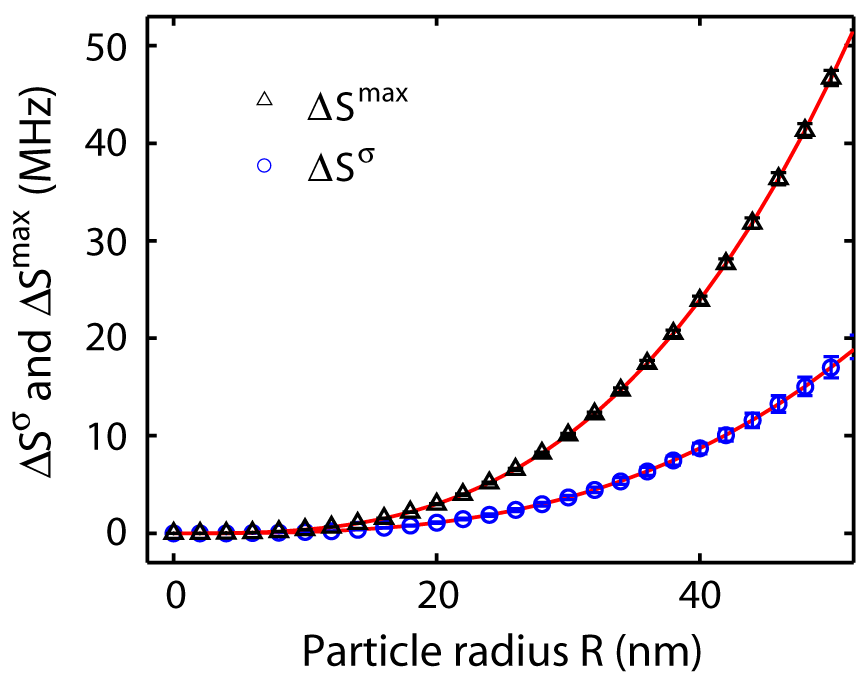}
  \caption{Expectations of $\Delta S^{\rm max}$ (black triangles) and $\Delta S^{\sigma}$  (blue circles) as a function of $R$ for PS particles. The number of particle binding events is $N$ = 200. The expectations (i.e., data points) are obtained from 10000 times repeated calculations. Red solid curves are polynomial fitting that scales with $R^3$.}
  \label{fig:SplittingStep_R}
\end{figure}

\begin{figure}[]
  \centering
  \includegraphics{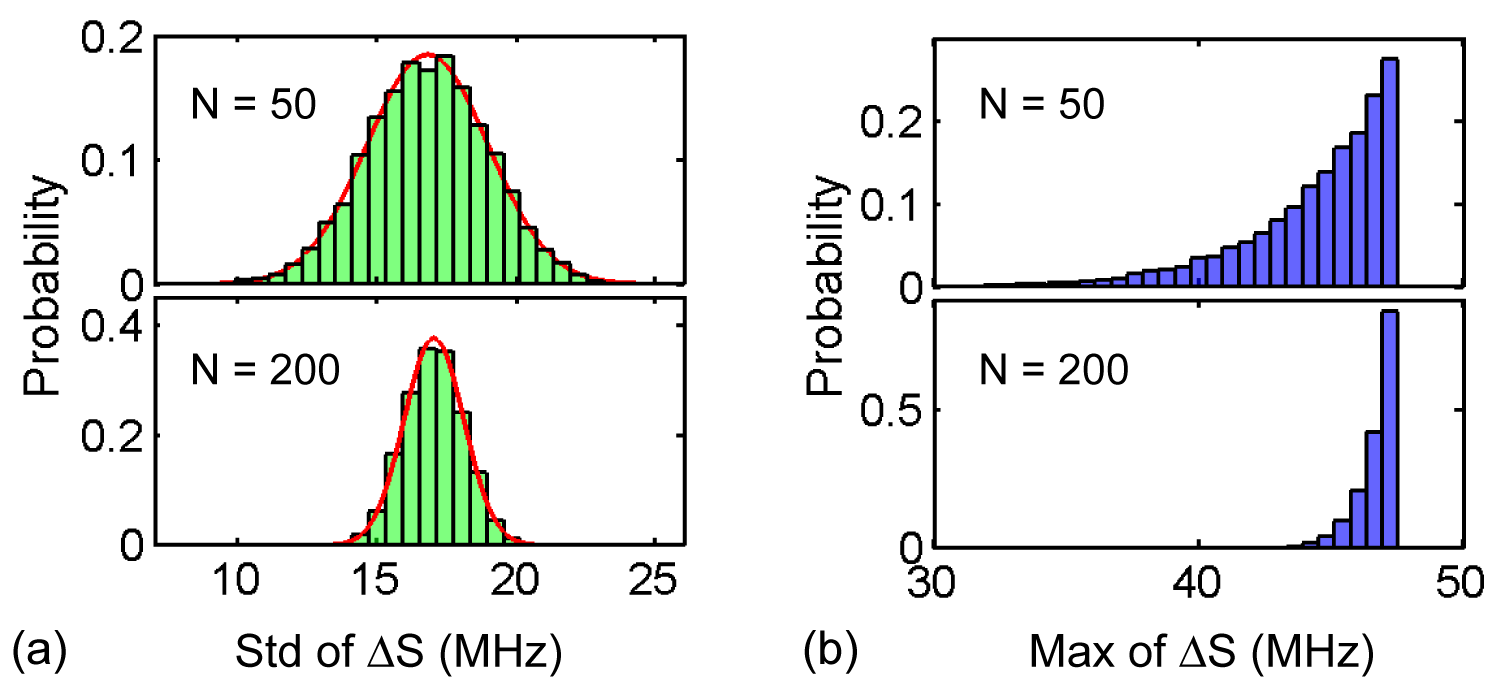}
  \caption{Distributions of $\Delta S^{\sigma}$ (a) and $\Delta S^{\rm max}$ (b) from 10000 times repeated simulations of PS particle ($R$ = 50 nm) deposition with different $N$. Solid red lines in (a) are Gaussian fittings.}
  \label{fig:Hist_Std_Max}
\end{figure}

\begin{figure}[]
  \centering
  \includegraphics{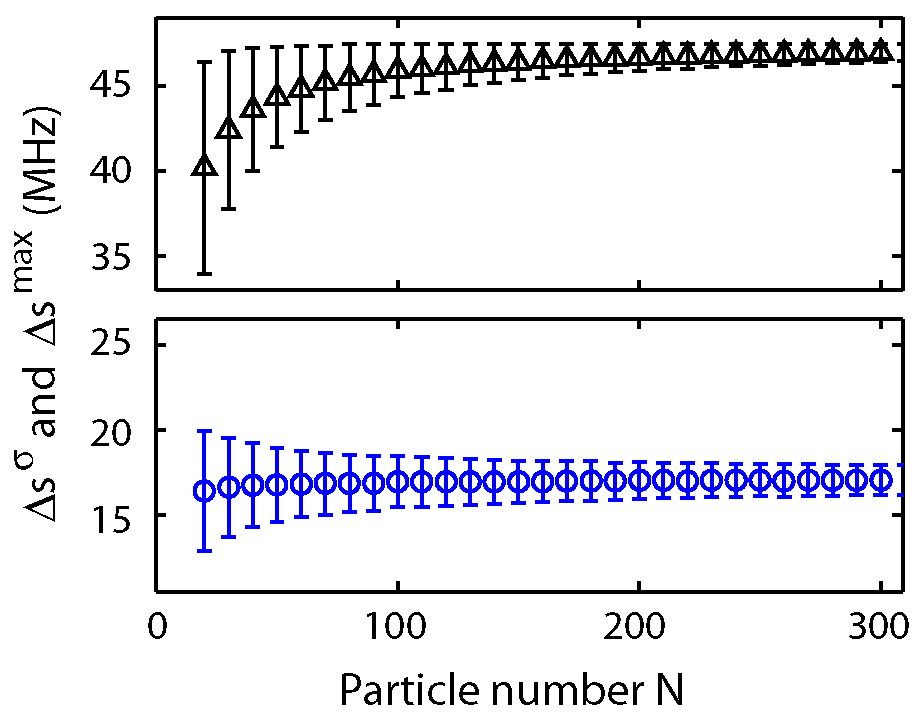}
  \caption{Mean (data points) and standard deviation (error bars) of $\Delta S^{\rm max}$ (black triangles) and $\Delta S^{\sigma}$ (blue circles) as a function of $N$ for PS particles of radius 50 nm. The results are obtained from 10000 repeated simulations.}
  \label{fig:SplittingStep_N}
\end{figure}

In each set of particle deposition experiment, $\Delta S^{\sigma}$ and $\Delta S^{\rm max}$ vary from their expected values due to the uncontrollable particle positions. We conduct numerical simulations to study quantitatively the dependence of the expectations of $\Delta S^{\sigma}$ and $\Delta S^{\rm max}$ on particle size. We perform the simulations as follows. $N$ particles of radius $R$ are randomly deposited into the mode volume, and the splitting changes $\Delta S_1\sim\Delta S_{\rm N}$ are calculated to obtain $\Delta S^{\sigma}$ and $\Delta S^{\rm max}$. This is repeated for $10000$ times to calculate their mean values which are plotted as a function of $R$ in Fig. \ref{fig:SplittingStep_R}. A linear dependence on $R^3$ is obtained for the expected values of $\Delta S^{\sigma}$ and $\Delta S^{\rm max}$. This linear dependence can be used to estimate the size of identical particles in an ensemble measurement. In real measurements, it is impractical to repeat experiments $10000$ times for each ensemble of particles to get the expected values of $\Delta S^{\sigma}$ and $\Delta S^{\rm max}$. Instead, one can perform only one ensemble measurement and use $\Delta S^{\sigma}$ and $\Delta S^{\rm max}$ values obtained in that specific measurement to estimate expectations. In such a case, the detected particle number $N$ in the ensemble is a crucial parameter determining the accuracy of the estimation. For small $N$, $\Delta S^{\sigma}$ and $\Delta S^{\rm max}$ in each measurement may vary a lot from their expectations, and thus leads to a large measurement error. However, errors can be reduced by increasing $N$, as shown in Fig. \ref{fig:Hist_Std_Max}. The standard deviations of $\Delta S^{\sigma}$ and $\Delta S^{\rm max}$ are smaller for larger $N$ (Fig. \ref{fig:SplittingStep_N}). The larger the $N$ is, the closer the estimated values to the real values are, and therefore the more accurate the particle measurement is.

In ensemble measurement of multiple-particle binding events, the particle polarizability can be estimated from the expression $\Delta S^{\rm max}=\alpha f_{\rm max}\!^2\omega_{\rm c}/V$ if the WGM field distribution and the mode volume are known. However, due to the many supported modes in a resonator, it is difficult to decide which mode is measured. In this case, reference measurements with particles of known size can be used to estimate the polarizability or size of unknown particles. For example, when two groups of particles of the same material but different size are deposited on a resonator, the ratio of the particle polarizability is equal to the ratio of $\Delta S^{\sigma}$ or $\Delta S^{\rm max}$ obtained for the two groups of particles. By comparing the ratio, we can eliminate the effects of the field distribution and the particle positions on the estimation. It should be noted that estimation using $\Delta S^{\rm max}$ gives only the lower limit for the particle polarizability because there is always a non-zero possibility that all the observed splitting changes are smaller than $\Delta S^{\rm max}$. Moreover, $\Delta S^{\rm max}$ is susceptible to perturbations due to contaminants on the resonator surface. In the measurements in Section \ref{sec:experiment}, we use $\Delta S^{\sigma}$ for particle size estimation.

\section{Measurements of nanoparticle size: Experimental results}\label{sec:experiment}

In a previous work, we have demonstrated detection and size measurement of nanoparticles down to $10~{\rm nm}$ in radius using a microlaser \cite{Lina_nature} obtained by optically pumping Erbium (Er$^{3+}$) or Ytterbium (Yb$^{3+}$) doped high-$Q$ microtoroidal resonators above the lasing threshold power levels. In the presence of mode splitting, single lasing frequency of the microlaser splits into two, which interfere and lead to a  beatnote signal when detected at a photodetector of sufficiently large bandwidth \cite{Lina_PRA,Lina_gain}. Changes in frequency splitting are thus translated to changes in the beatnote frequency which is processed as the sensing signal. The ultra-narrow laser linewidth allows detection of small particles which will go undetected if a passive resonator was used. However, it is difficult, if not impossible, to measure the linewidth difference between the split lasing lines \cite{TLuVahala}. This, in turn, makes it difficult to extract the size information of the detected particle at a single-shot measurement. Thus, ensemble measurements of particles with identical size should be performed to assign an average size or polarizability to the detected particle ensemble. In this section, we present size measurement results for sodium chloride (NaCl) and gold nanoparticles using Er$^{3+}$ doped silica microtoroid lasers \cite{Yang_laser,Lina_SP}, and study the dependence of size estimation accuracy on the number of particles in the ensemble.

\begin{figure}[]
\centering
\includegraphics{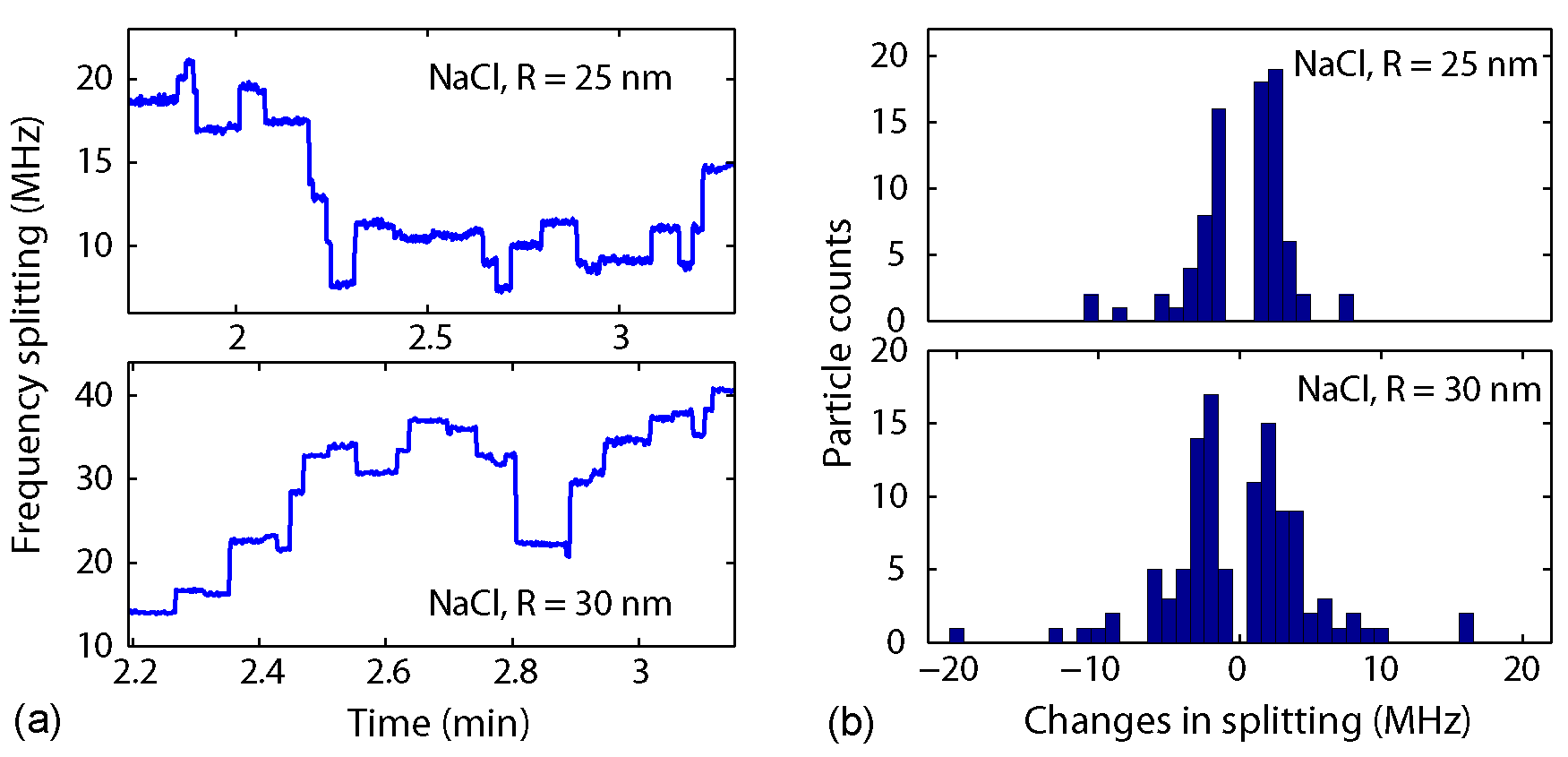}
\caption{(a) Real-time record of frequency splitting as NaCl particles consecutively bind on a microtoroidal laser. (b) Histograms of the changes in frequency splitting induced by NaCl particles of different sizes. Particle radius is labeled in the plot. Total particle number in the histograms is 81 for 25nm-particle and 111 for 30nm-particle.}
 \label{fig:NaCl}
\end{figure}

Figure \ref{fig:NaCl}(a) depicts the frequency of the beat signal (i.e., frequency splitting) observed as individual NaCl nanoparticles ($R=25 {\rm ~nm}$ and $R=30 {\rm ~nm}$) are adsorbed into the mode volume of the microlaser. In Fig. \ref{fig:NaCl}(a), each discrete change in frequency splitting indicates adsorption of a particle. The statistical distribution of the changes in the splitting is plotted in Fig. \ref{fig:NaCl}(b). We have rejected splitting  changes which lie within the noise level of beat-note signal. Thus, the histograms in Fig. \ref{fig:NaCl}(b) depicts gaps around 0 value. As expected, larger particles lead to broader distributions. By setting one group of particles as reference, size of the other particle group is estimated by taking the ratio of $\Delta S^{\sigma}$ from the measured dataset. Due to the missing information near 0 in the histograms, $\Delta S^{\sigma}$ is obtained at different threshold values to get the particle size. Detailed explanations can be found in the Supplementary information of \cite{Lina_nature}. For the ensemble of particles with $R=30~{\rm nm}$ ($R=25~{\rm nm}$), we estimate the size as $R=30.42~{\rm nm}$ ($R=24.81~{\rm nm}$) when the measurement results with particle ensemble of $R=25~{\rm nm}$ ($R=30~{\rm nm}$) are used as reference.

The estimation results agree with the size information provided by the manufacturer, but this single estimation result does not tell us how accurate it is as an estimate of the true value. To determine the accuracy of size estimation, we use {\it bootstrap method} to obtain the confidence interval of the estimate \cite{bootstrap1,bootstrap2}. In the bootstrap approach, the measured original dataset is randomly resampled to form a new dataset having the same length as the original dataset. Each of the resampled data is obtained from random sampling and replacement of the original data points. For each resampled dataset, we calculate the particle size. This resampling process is done $1000$ times giving us a collection of $1000$ size estimations. The distribution of these estimations approximates the distribution of the actual particle size. The confidence interval can thus be obtained by using the appropriate upper and lower percentages of the distribution. Using this method, $95\%$ confidence intervals for measurements in Fig. \ref{fig:NaCl} are found as ($21.38 {\rm nm}$, $29.49 {\rm nm}$)  and ($25.41 {\rm nm}$, $35.05 {\rm nm}$) for two different particle sizes.

\begin{figure}[]
  \centering
  \includegraphics{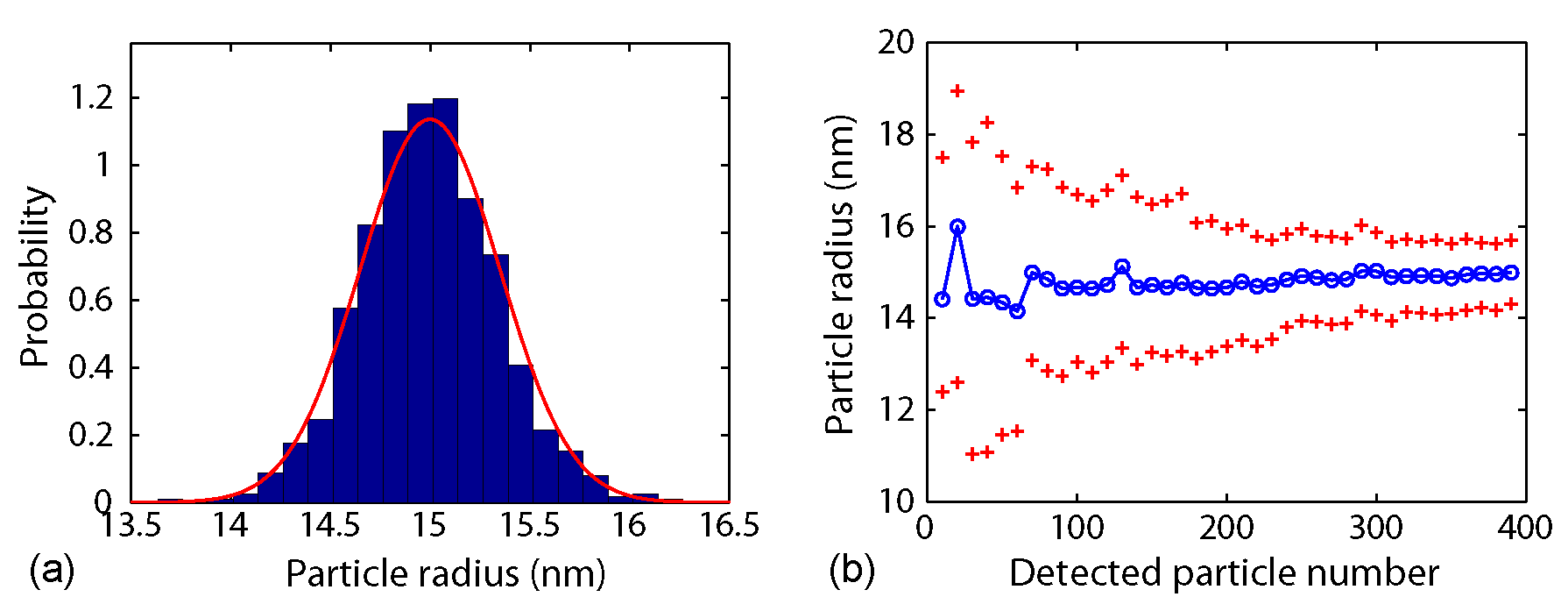}
  \caption{Size estimation of gold nanoparticles. Particles with 25nm-radius are taken as reference to measure the size of another group of gold particles which have radius of 15 nm. (a) Distribution of size estimation using bootstrap method of 1000 resamples. Red curve is Gaussian fitting. In the measurement, we detected around 400 reference particles and 400 measured particles. (b) Size estimation as a function of detected particle number for both reference and measured particles. Blue circles and red crosses are the mean and 95\% confidence interval of size estimation obtained from bootstrap method of 1000 resampling.}
  \label{fig:Au30nm}
\end{figure}

One can reduce the estimation error by increasing the number of particles that are detected. We performed experiments with gold nanoparticles to verify it. Using a single microtoroidal laser, we detected around $400$ particles of $R=15~{\rm nm}$ and $400$ reference particles of $R=25~{\rm nm}$. Distribution of the estimated size using bootstrap method is shown in Fig. \ref{fig:Au30nm}(a) with mean value of $14.97~{\rm nm}$. To study the effect of the number of detected particles on the accuracy of size estimation, we chose the first $N$ ($1\leq N\leq400$) data points from the original measured data set as the new data set to estimate the particle size and its $95\%$ confidence interval. The results as a function of $N$ are plotted in Fig. \ref{fig:Au30nm}(b). Obviously, the confidence interval decreases with increasing particle number, suggesting a higher accuracy. For particle number of $N=250$, the mean value of estimated size is $R=14.92~{\rm nm}$ with the confidence interval of $[13.94~{\rm nm}-15.94~ {\rm nm}]$ which is close to the data provided by the manufacture (mean $15.15~{\rm nm}$ with the coefficient of variance of $<8\%$). These results imply that the proposed size measurement method does not necessarily require that all the particles in the ensemble have exactly the same size, although for the numerical simulations we considered the same size. Ensembles of particles of the same material and shape but not exactly the same size (i.e., size of the particles in the ensemble obeys a statistical distribution with a mean and standard deviation) can also be characterized with the proposed method.

Finally, we would like to discuss the similarities and differences of size measurement method proposed and demonstrated in this work and that used in the reactive-sensing scheme by Vollmer {\it et al.} \cite{Vollmer_PNAS}, who have provided an analytical expression which relates the resonance shift to the particle size and the size of the microsphere resonator. In the reactive-sensing scheme with passive resonators, with the help of an analytical expression, an average size is assigned to an ensemble of identical particles using the observed maximum frequency shift after a number of particles bind to the resonator. In this method only the maximum frequency shift is used but the distribution of the all observed frequency shifts is ignored. In such estimations based on only one value chosen from the statistical distributions of the events, one may expect to have larger errors. For example, it will be always questionable whether the maximum shift that one observes after, say $N$ binding events, is still the maximum shift after $(N+1)$-th binding events, and how one can be sure that the maximum observed shift is due to binding of a single particle of interest but not due to the binding of aggregated particles or some other larger particles present in the solution. On the other hand, making use of the whole set of obtained results and their statistical distribution will help reduce false measurements due to outliers. Therefore, we expect that the method we propose in this study will provide a better size estimation.

The method proposed in this paper can be directly applied to reactive sensing schemes too. We should emphasize here that in particle detection and measurement using mode splitting in passive WGM resonators, we do not need the method proposed here. In this case, as we have shown previously \cite{Zhu_OE_MultiVirus,Jiangang_nature}, we can directly estimate the size of each particle by comparing the mode splitting spectra (changes in the frequencies and linewidths of the split modes) just before and after a single particle binding. We proposed this new method of size estimation for WGM microcavity-laser-based sensing, because in this case the ultranarrow laser linewidths are difficult to measure accurately. This method can also be used for measuring small nanoparticles which do not induce sufficiently high dissipation that can create a measurable amount of changes in linewidths. Since linewidth information is missing, we cannot measure the size of each detected particle but instead we give an average size for the ensemble using the statistics of the changes in mode splitting. In principle, reactive-sensing scheme can be realized using microlasers, too, by monitoring the shifts in the lasing frequency. However, the laser frequency shift induced by small nanoparticles is so small that one cannot use commercially available optical spectrum analyzers to monitor the shift induced by a single nanoparticle binding, and novel measurements techniques are required. On the other hand, mode splitting method is easy to use for microlaser-based sensing because the signal at the detector is a beatnote signal whose frequency corresponds to the amount of mode splitting. Microlaser-based sensing methods will certainly benefit from research targeting the detection of ultranarrow linewidths and the changes in linewidths.

\section{Conclusions}
In this paper, we have studied numerically and experimentally the statistical properties of MPIFS. We have showed by simulations and experiments that the difficulty of precisely controlling the positions of the adsorbed particles within the mode volume leads to decrease or increase in the amount of frequency splitting with varying step heights as the particles are adsorbed one-by-one onto the resonator. Despite this randomness in frequency splitting for each individual binding events, statistical analysis shows that the expected value of frequency splitting increases linearly with the square root of particle number. We have demonstrated that statistics of changes in frequency splitting can be used to extract the  information of particle polarizability and hence the particle size if the refractive index of the particles is known. Although our experiments are performed with NaCl and Au nanoparticles and numerical simulations consider PS and Au nanoparticles, our results and the proposed size measurement method are valid for all types of particles. However, as mentioned before, this study does not consider plasmonic effects. This statistical analysis based size and polarizability estimation method can be used in both the resonance-shift based detection (reactive sensing) and laser frequency splitting techniques where the amount of spectral shift and change in splitting frequency are measured. A possible application area of the proposed method is the characterization of nanoparticle generators and sources.

\section*{Acknowledgement}
This work was supported by NSF under Grant  number 0954941 and the U. S. Army Research Office under grant number  W911NF-12-1-0026.

\newpage
\section*{FIGURES}

\begin{figure}[]
  \centering
  \includegraphics{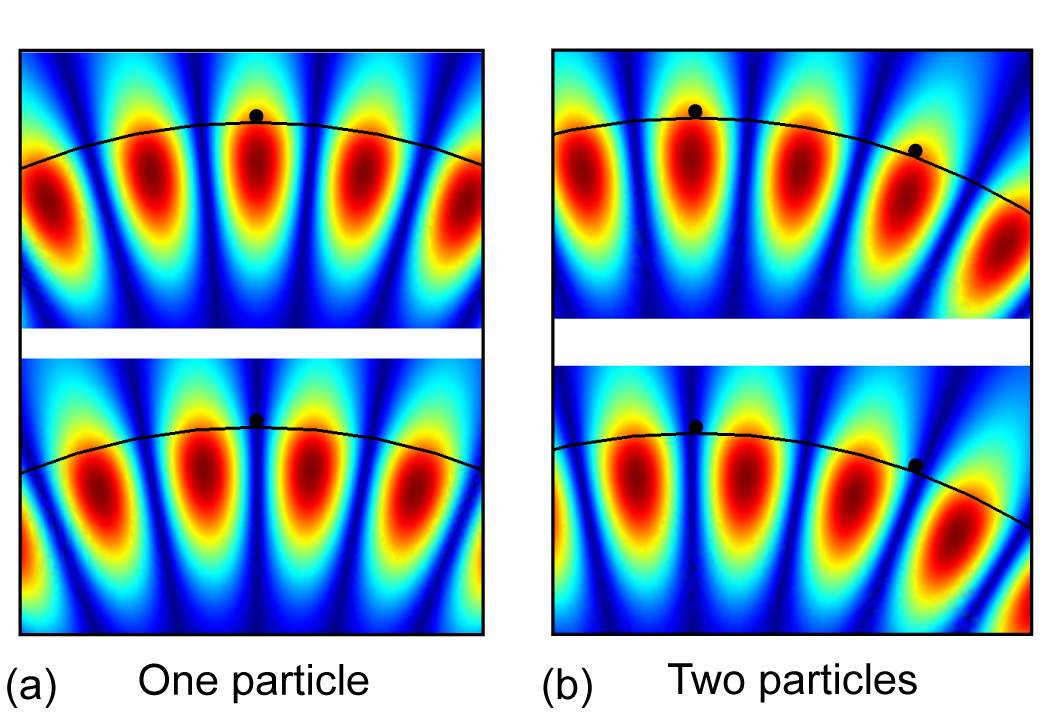}
  \caption{Results of numerical simulations depicting the distributions of the two standing wave modes in response to one single particle (a) and two particles (b) on a resonator. The black curves denote the resonator surface and the black dots represent particles. }
  \label{fig:Toroid-particles}
\end{figure}


\end{document}